\documentclass[3p,10pt,sort&compress, colorlinks,linkcolor=blue,square, citecolor=blue]{elsarticlex}

\usepackage[utf8]{inputenc}

\usepackage{amssymb}
\usepackage{amsmath,mathrsfs}
\usepackage{color}

\usepackage{extarrows}

\usepackage{mathrsfs}
\usepackage{empheq}

\usepackage{booktabs}

\usepackage[sc]{mathpazo}
\usepackage{bm}

\usepackage{color,soul}

\usepackage{extarrows}

\usepackage{verbatim}

\usepackage{lineno}

\usepackage{siunitx}

\usepackage{stfloats}

\usepackage{bm}

\usepackage{graphicx}

\usepackage{caption}
\usepackage{subcaption}

\usepackage{makecell}

\usepackage{stmaryrd}

\usepackage[super]{nth}

\usepackage{mhchem}

\usepackage[toc,page]{appendix}

\usepackage{bm}

\usepackage{float}

\usepackage{empheq}

\usepackage{enumerate}

\usepackage{tabularx}

\usepackage[toc,page]{appendix}
\usepackage[colorlinks,linkcolor=blue,anchorcolor=green,citecolor=blue]{hyperref}

\newcommand{\secref}[1]{Sec.~\ref{#1}}

\newcommand{\figref}[1]{Fig.~\ref{#1}}
\newcommand{\subfigref}[2]{Fig.~\ref{#1}\ce{#2}}
\newcommand{\subsfigref}[3]{Figs.~\ref{#1}\ce{#2} and \ref{#1}\ce{#3}}
\newcommand{\subssfigref}[3]{Figs.~\ref{#1}\ce{#2}-\ref{#1}\ce{#3}}
\newcommand{\tabref}[1]{Table.~\ref{#1}}
\renewcommand{\eqref}[1]{Eq.~$($\ref{#1}$)$}

\begin{document}
	
	%opening
	
	\begin{frontmatter}
		\title{Microstructural influence on hysteresis and eddy current losses of additively manufactured electrical steel: A multiphysical approach}

			\author[els]{Patrick Kühn}

            \author[els]{Yangyiwei Yang\corref{cor1}}
			\ead{yangyiwei.yang@mfm.tu-darmstadt.de}

            \author[els]{Guanyu Chen}
			
            \author[alt]{Shanelle N. Foster}

            \author[sad,sad2]{Herbert Egger}

            \author[els]{Bai-Xiang Xu\corref{cor1}}
			\ead{xu@mfm.tu-darmstadt.de}

			\address[els]{Mechanics of Functional Materials Division, Institute of Materials Science, Technische Universität \" at Darmstadt, Darmstadt 64287, Germany}
	
			\address[alt]{Department of Electrical and Computer Engineering, Michigan State University, 48824, MI, United States}

            \address[sad]{Johann Radon Institute for Computational and Applied Mathematics, Linz, Austria}
            \address[sad2]{Institute for Numerical Mathematics, Johannes Kepler University, Linz, Austria}
			
			\cortext[cor1]{Corresponding author}

		%%\author[mymainaddress,mysecondaryaddress]{Elsevier Inc}
		%%\ead[url]{www.elsevier.com}

		\begin{abstract}
Improving the efficiency of electrical machines requires fundamental knowledge of the mechanisms behind magnetic and eddy current losses of the magnetic core materials, with Fe-Si alloy as a prototype. The microstructure of the materials intrinsically influences these losses. In this work, we calculated the hysteresis and eddy current losses of the additively manufactured electrical steel using a multiphysical approach, which consists of the demagnetization simulation based on the Landau-Lifshitz theory and computational homogenization based on the magneto-quasi-static (MQS) Maxwells equation. 
To demonstrate this method, the microstructure samples were digitized and generated based on experimental characterization of the additively manufactured Fe-Si-B steel under different Boron compositions. By performing parameter research on a series of digital microstructures, effects of average grain size and grain boundary (GB) phase thickness on the hysteresis loss and eddy current loss were unveiled. An average grain size around 120 \si{\micro m} has the lowest hysteresis loss, although the eddy current loss increases with the grain size. Increasing GB-phase thickness helps reduce both losses.
Results indicate the potential to decrease loss of magnetic core materials by microstructure optimization. 
		\end{abstract}
		
		
	\end{frontmatter}

\def\iu{\mathsf{j}}
\def\fsb{\ce{FeSiB}}
\def\fesi{\ce{FeSi}}
\def\ra{\rightarrow}
\def\mumax{\texttt{mumax}^3}

%%% volume
\def\vol{V}
\def\BH{B\text{-}H}
\def\dV{\text{d}V}
%%%%%%%%%%%%%%%%%%%%%%%%%%%%%%%%%%%%%%%%%%%
%% Operator
%%%%%%%%%%%%%%%%%%%%%%%%%%%%%%%%%%%%%%%%%%%
% for full derivation
\newcommand*{\diff}{\mathop{}\!\mathrm{d}}
\newcommand*{\Diff}{\mathop{}\!\mathrm{D}}

\newcommand*{\pd}[2]{\mathop{}\!\frac{\partial #1}{\partial #2}}
\newcommand*{\varid}[2]{\mathop{}\!\frac{\delta #1}{\delta #2}}
\newcommand*{\fed}[1]{f_\mathrm{#1}}

%%%%%%%%%%%%%%%%%%%%%%%%%%%%%%%%%%%%%%%%%%%
%% Thermodynamic
%%%%%%%%%%%%%%%%%%%%%%%%%%%%%%%%%%%%%%%%%%%

%% functional
\def\fedf{\mathscr{F}}

%%%%%%%%%%%%%%%%%%%%%%%%%%%%%%%%%%%%%%%%%%%
%% Statistics
%%%%%%%%%%%%%%%%%%%%%%%%%%%%%%%%%%%%%%%%%%%

\def\Rsq{\mathrm{R}^2}
\def\mse{\mathrm{MSE}}

%%%%%%%%%%%%%%%%%%%%%%%%%%%%%%%%%%%%%%%%%%%
%% Notations
%%%%%%%%%%%%%%%%%%%%%%%%%%%%%%%%%%%%%%%%%%%
% sci notation
\newcommand*{\E}[1]{\mathop{}\!\times 10^{#1}}

\def\deg{^{\circ}}

%% Vol. Frac.
\def\vf15{\varphi_{S}}
\def\vfz{\varphi_\mathrm{Z}}

%%%%%%%%%%%%%%%%%%%%%%%%%%%%%%%%%%%%%%%%%%%
%% structural
%%%%%%%%%%%%%%%%%%%%%%%%%%%%%%%%%%%%%%%%%%%
\def\ws{w_\mathrm{S}}
\def\wz{w_\mathrm{Z}}

%%%%%%%%%%%%%%%%%%%%%%%%%%%%%%%%%%%%%%%%%%%
%% Magnetics
%%%%%%%%%%%%%%%%%%%%%%%%%%%%%%%%%%%%%%%%%%%
\def\Ku{K_\mathrm{u}}
\def\Ms{M_\mathrm{s}}
\def\Bs{B_\mathrm{s}}
\def\mv{\mathbf{m}}
\def\pos{\mathbf{r}}
\def\mag{\mathbf{m}}
\def\H{\mathbf{H}}
\def\Hx{\mathbf{H}_\mathrm{ext}}
\def\Hdm{\mathbf{H}_\mathrm{dm}}
\def\Hc{\mathbf{H}_\mathrm{c}}
%%params
\def\Ae{\kappa_\mathrm{m}}
\def\sigb{\sigma_\mathrm{B}}
\def\sign{\sigma_\mathrm{B}}
\def\ldw{l_\mathrm{B}}

%% easyaxis
\def\eau{\mathbf{u}}

\def\Hc{H_\mathrm{c}}
\def\Br{B_\mathrm{r}}
\def\Hp{H_\mathrm{p}}
\def\Hani{H_\mathrm{ani}}

%% pining fraction
\def\Pp{\Phi_\mathrm{p}}

\def\Hv{\mathbf{H}}
\def\Hvext{\mathbf{H}_\mathrm{ext}}
\def\Bv{\mathbf{B}}
\def\Mv{\mathbf{M}}

\def\Hdm{\mathbf{H}_\mathrm{dm}}

%% orientation angle
\def\oriang{\vartheta}

\def\bHc{\bar{H}_\mathrm{c}}

%%variation
\def\delFdelm{\frac{\delta \mathcal{F}}{\delta \mathbf{m}}}

\def\ngl{l_\mathrm{nc}}
\def\fv{\varphi_\mathrm{nc}}

\def\gl{d_\mathrm{G}}
\def\gbw{\ell_\mathrm{GB}}

\def\eff{\mathrm{eff}}

%%%% Loss
\def\PH{P_\mathrm{H}}
\def\PE{P_\mathrm{E}}
\def\WH{W_\text{H}}
\def\WE{W_\text{E}}

%%% vector things
\def\mvp{\boldsymbol{A}}
\def\cuv{\hat{\boldsymbol{e}}_i}
\def\XSi{X_{\ce{Si}}}

%%% electrical
\def\Ev{\boldsymbol{E}}
\def\Jv{\boldsymbol{J}}
\def\econd{\boldsymbol{\sigma}}

\section{Introduction}

Electrical steel, generally refers to the $\fesi$ alloy system, is one type of soft magnets that has been widely employed in electronics since their development in the early 19th century. Electrical steels are iron silicon alloys with between $0$ and $6.5~$ wt\% silicon. As of now, electrical steel is by far the most widely used soft magnetic material. 
Their importance only grows with today's trend of moving away from a fossil-fuel based energy generation and conversion, as they are a key component of rotating machines. These rotating machines are employed on a wide scale from generators and transportation sector motors down to the smallest electric machines, such as an electric toothbrush\cite{moses1990electrical}. In these applications the electrical steel usually is subjected to AC conditions, which in turn causes energy losses, either due to hysteresis or eddy currents. For developed countries some estimates put the magnitude of these losses between $5$ and $10~$ \% of all power generated. Therefore reducing these losses has been an active area of research for the last century. While a significant reduction of losses could already be achieved. Additional research may be able to improve this even further\cite{moses2012energy}.

Magnetic properties of materials are influenced by their microstructure and texture, which in turn originates from mechanical and thermal processing of the material.  For instance electrical steel with the Goss-texture is produced by a combination of rolling and recrystallization annealing. It shows strong magnetic anisotropy, making them a good choice for transformer cores with minimal hysteresis loss\cite{jain2022origin}. Other processing steps, such as cutting or punching can locally change the microstructure and texture and introduce stresses in the material \cite{gschwentner2023determination, weiss2018impact}. Conventional production of electrical steels involves rolling requiring good workability of the material, but at a Si content of $4$ wt\% FeSi transitions from the B$_2$ phase (bcc) to the Fe$_3$Si (D0$_3$ crystal structure)\cite{ustinovshchikov2005ordering}. As the D0$_3$ phase causes the material behavior to become hard and brittle, processing via rolling becomes a challenge, especially as increasing the Si content increases electric resistivity and magnetic permeability.

A novel way to process soft magnetic materials is additive manufacturing (AM), circumventing the rolling completely. Reviewed works mostly focused on Fe-Ni produced via powder bed fusion\cite{yang2023npj, schonrath2019additive, zhang2012studies}, but current works have extended this to the screen printing of electrical steel\cite{mix2023additive}. Another route of processing electrical steel is binder-jet printing (BJP). With BJP a 3D-body can be produced from a powder material, which is joined layer by layer with the aid of a polymeric binder liquid\cite{mostafaei2021binder}. Other than just providing a new processing route this method has been reported to improve the overall magnetic properties. This includes a lowered intrinsic coercivity\cite{pham2020reduction, kumari2023improving}, reduced hysteresis loss\cite{pham2020reduction} and reduced eddy current losses\cite{islam2022eddy}. These effects could be observed using Boron as an additive. Kumari \textit{et al.} \cite{kumari2023improving} showed distinct impacts of Boron addition on the microstructure, it increased the grain size and lead to the evolution of a lamellar grain boundary(GB)-phase, consisting of Fe$_2$B and $\fesi$. This in turn reduced the intrinsic coercivity, but also lead to a slight decrease in saturation flux density, though deemed negligible; furthermore, the permeability was improved. Nevertheless, the underlying mechanisms changing the magnetic properties, electric properties and losses remain unclear.

In this work we investigate the mechanisms behind this changed behavior. Microstructures are reconstructed digitally, both through a synthetic descriptor based approach and a direct digitization approach. Based on these reconstructions we perform demagnetization simulations in order to analyze how the GB-phase and modified grain size influence the domain structure, coercivity, remanence and hysteresis loss. This data is then extended with a series of simulations, based on synthetic reconstructions within the experimentally observed ranges of structure parameters. Furthermore we calculate the effective electric properties of the different microstructure and analyze their behavior in response to changes of the structural parameters. We present how currents flow through the different microstructures allowing us to identify bottlenecks and preferred pathways. These effective properties are then used to calculate the eddy current losses for different microstructures. It is our belief that this work will further the understanding of the underlying mechanisms in soft magnetic materials and open up improved ways to design soft magnetic materials.

\section{Multiphysical approach}

\subsection{Microstructure reconstruction and digitization}\label{sec:wf}

\begin{figure*}
	\centering
 	\includegraphics[width=18cm]{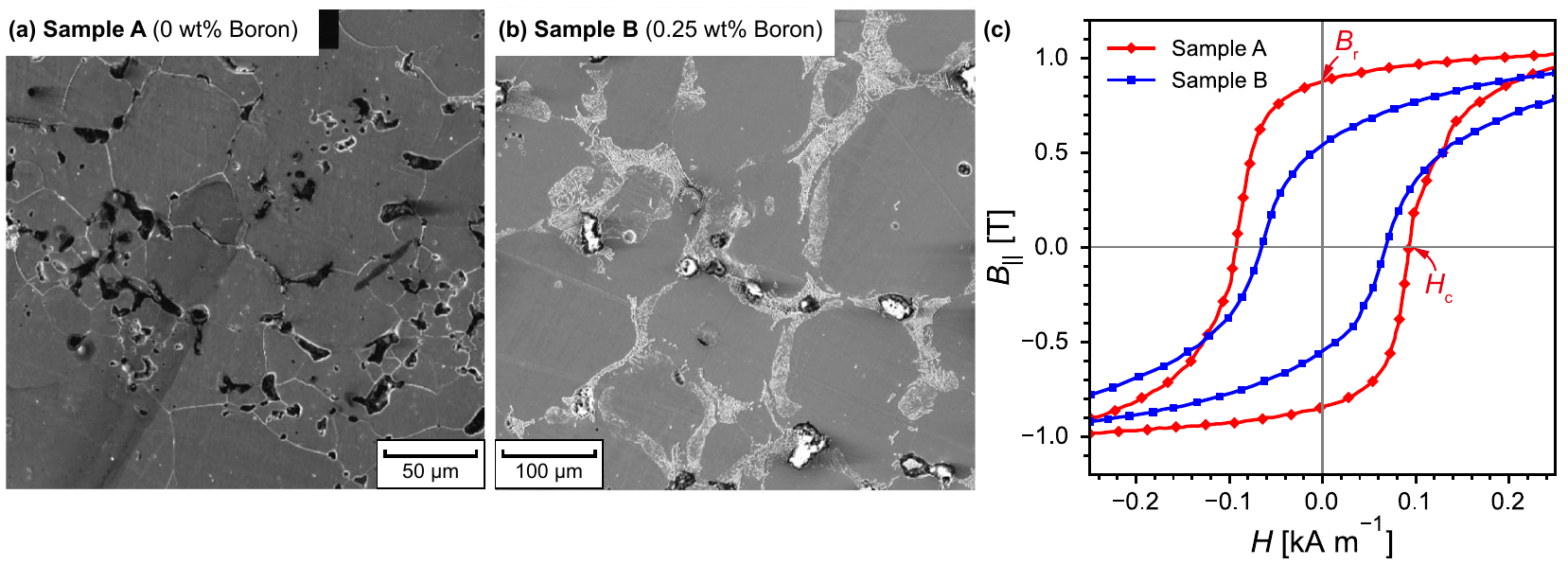}
	\caption{SEM photomicrographs of (a) sample A and (b) sample B. (c) shows the quasi-static hysteresis loop of sample A and B. The magnetic coercivity $\Hc$ and the magnetic remanence $\Br$ are also denoted.}
	\label{fig:sc1}
\end{figure*}

Two samples from Kumari \textit{et al.} \cite{kumari2023improving} were selected for digital reconstruction. These represent their samples with the highest and lowest coercivity, correspondingly, as presented in Fig. \ref{fig:sc1}.
The samples are also distinguished by the chemical composition, i.e., sample A contains 3 wt\% Si and Boron-free, and sample B contains 5 wt\% Si and 0.25 wt\% Boron.  
Fig. \ref{fig:sc1} (a) and (b) present the microstructures of sample A and B together with the corresponding hysteresis cycles in \figref{fig:sc1} (c). Several features can be directly identified from the corresponding microstructures. Notably, sample B presents a greatly enlarged grain size, reduced porosity, and the grain boundary region filled with Fe/Fe$_2$B, whereas sample A presents a relatively smaller grain, more pores, and thin grain boundaries. As indidcated by the hysteresis measurements, sample A presents relatively larger magnetic coercivity ($\Hc$) and remanence ($\Br$) compared to sample B. There are already existing works investigating the trend of these magnetic properties under varying microstructural descriptors. For instance, Igarashi \textit{et al.} demonstrated a negative correlation between $\Br$ and porosity, while $\Hc$ presents almost no dependency on porosity \cite{igarashi1977effects}. On the other hand, $\Hc$ is almost linearly related to the reciprocal of the average grain size, while $\Br$ presents almost no dependency on grain size. Nonetheless, the reviewed work fails to explain the comparably larger $\Hc$ and $\Br$ in sample A, possibly due to the lack of consideration for the morphology and properties of the existing GB-phase. This motivates us to start this work by developing pipelines for reconstructing/digitizing the experimentally characterized microstructures.

\begin{figure*}
	\centering
 	\includegraphics[width=18cm]{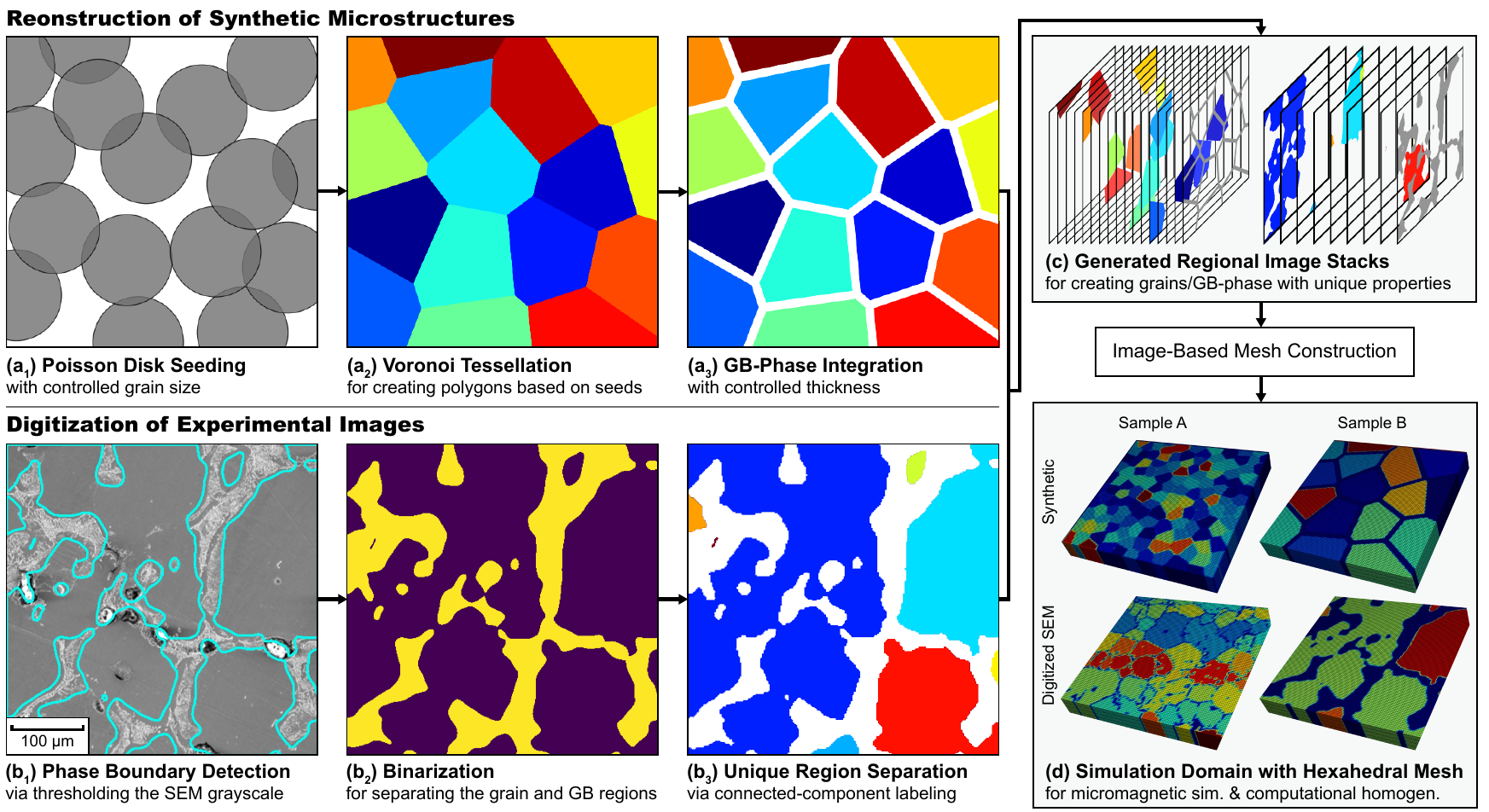}
	\caption{Schematic of the workflows for microstructure generation, with the synthetic approach through poisson seeding the the grain positions (a$_1$), grain generation through Voronoi tessellation (a$_2$) and integration of the GB-phase (a$_3$). (b) shows the direct digitization, from edge detection and conversion of a SEM photomicrograph to grayscale(b$_1$), followed by binarization of the image, separating the grains and grain boundary regions (b$_2$) and the identification of separated regions (b$_3$). Both cases are used to generate a stack of images (c) which are then translated into microstructures (d).  }
	\label{fig:sc}
\end{figure*}

Two workflows are employed in this work to investigate {the influence of the grain size and the GB-phase on the magnetic hysteresis and corresponding losses}. 
The first workflow reconstructs synthetic polycrystalline microstructures by specifying statistic features (like average grain size)  from the experimental characterization; see \subssfigref{fig:sc}{a1}{a3}. It starts from generating random-labeled seeds with controlled diameter (\subfigref{fig:sc}{a1}). This is based on the fast Poisson disk sampling algorithm to have a reasonable distance among seeds and avoid significant overlap \cite{bridson2007fast}.  
Voronoi tessellation is then applied to create polygonal cells that represent grains based on sample seeds (\subfigref{fig:sc}{a2}). The average grain size $\gl$ can be then statistically approximated by the seed diameter. Next, the GB-phase is integrated into the microstructure via illustrating an outline of the Voronoi cells with an evenly controlled thickness $\gbw$ (\subfigref{fig:sc}{a3}). For the microstructure without GB-phase, $\gbw=0$. The generated synthetic polycrystalline microstructures with the GB-phase are then split into an image stack based on their labels, as the indicated by the colors shown in \subsfigref{fig:sc}{a3}{d}. To be specific, the grains (Voronoi cells) inherit their labels as $1\sim N$ from the random-labeled seeds, while GB-phase is labeled as zero by default. Finally, the images from the stack are used to construct regions in the simulation domains assigned unique properties. These hexahedral-mesh domains can be used for simulations based on either finite difference (FDM) or finite element (FEM) methods. This workflow can deliver microstructures with controlled parameters, notably $\gl$ as well as $\gbw$, and is suitable for subsequent parameter research. However, the generated microstructures are ideal compared to those from experimental characterization. Many features that might influence the magnetic hysteresis, such as the stochastic morphology and the inhomogeneous thickness of the GB-phase, are also excluded. 

To supplement the first one, the second workflow directly digitizes the experimental SEM images and converts them into the simulation domains. First, the phase boundary is detected with a specified gray-scale threshold on an SEM image that has undergone Gaussian smoothing to remove intricate details \cite{fathidoost2024thermal} (\subfigref{fig:sc}{b1}). Then, the image is binarized based on detected boundaries, resulting in one with regions representing grains and GB-phase separated (\subfigref{fig:sc}{b2}). Next, to further identify the disconnected regions as unique grains and GB-phase for the subsequent property assignment, connected-component labeling algorithm \cite{dillencourt1992general} is applied on the binary images (\subfigref{fig:sc}{b3}). Up to this step, an  SEM image has been digitized in the form of an array containing labels $1\sim N$, which has the same data-structure as one generated synthetically. Therefore, the remaining steps are identical to the first workflow, as shown in \subsfigref{fig:sc}{c}{d}. This workflow can deliver microstructures in a form closest to the experimental observation, but it has limited adjustability compared to the first workflow and its generated microstructures can be vastly influenced by the choice of the gray-scale threshold during binarization. Notably, the underlying pores cannot be well distinguished with grains as both of them have higher gray-scale value comparing to the GB-phase. Future works will expand upon this method, by introducing improved phase, phase-boundary and pore detection.

\subsection{Hysteresis loss from Landau-Liftshitz-based demagnetization simulation}\label{sec:mmhl}

Hysteresis loss is the energy loss in the magnetic cores during one cycle of the magnetization-demagnetization process (i.e., a complete hysteresis loop). It is the main core loss at low operation frequencies \cite{shokrollahi2007soft}. In the proposed approach, the hysteresis loop is obtained by performing demagnetization simulations based on Landau-Liftshitz theory. The theory employs a semi-classical approach to represent magnetization. As a consequence, a position-dependent vector $\mv (\pos)$ is introduced to physically interpret the distribution of magnetization as the mean field of the local atom magnetic moments inside a ferromagnetic material. Meanwhile, it requires the simulation domain to be sufficiently small in scale to resolve the magnetization transition across the domain wall. In this regard, the magnetic free energy functional of a simulation domain with a volume of $\vol$ can be formulated as
  
\begin{equation}
\fedf=\int_\vol[\underbrace{\fed{ani}+\fed{ms}+\fed{zm}}_{\fed{loc}}+\fed{grad}]\diff \vol,
    \label{eq:fmag}
\end{equation}
with
\begin{align*}
&\fed{grad}(\nabla\mv) = \Ae \|\nabla\mv\|^2, \\
&\fed{ani}(\mathbf{m}) = - K_{\mathrm{u}}\left(\eau\cdot\mv\right)^{2}, \\
&\fed{ms}(\mathbf{m}) = -\frac{1}{2}\mu_0 M_\mathrm{s}\mv\cdot\Hv_\mathrm{dm}, \\
&\fed{zm}(\mathbf{m}, \Hx) = -\mu_0M_\mathrm{s}\mv\cdot\Hx,
\end{align*}
%Here, $f_\mathrm{ex}$ is the exchange contribution, recapitulating the parallel-aligning tendency among neighboring magnetic moments due to the Heisenberg exchange interaction. The norm $\| \nabla\mv\|$ here represents $\sum_j|\nabla m_j|^2$ with $j=x,y,z$ and $\mv=[m_x, m_y,m_z]$.
where $f_\mathrm{grad}$ represents the energy penalty from the domain wall represented by the continuous vectoral field of magnetization. Here $\| \nabla\mv\|$ represents $\sum_j|\nabla m_j|^2$ with $j=x,y,z$ with $\mv=[m_x, m_y,m_z]$, and $\Ae$ is the gradient constant. Adopting an interpretation from micromagnetics, it states the contribution that originates from the Heisenberg exchange interaction, driving neighboring spins into parallel alignment. In this regard, $\Ae$ is also known as the exchange stiffness coefficient. 
$f_\mathrm{loc}$ represents the local energy density landscape in the simulation domain, presenting multiple local minima that denote quasi-static configuration magnetization under a certain condition. By micromagnetics, $f_\mathrm{loc}$ is generally considered as a sum of the anisotropic term $\fed{ani}$ counting the energy due to the mis-match between magnetization and a crystallographic easy-axis $\eau$, the magnetostatic term $\fed{ms}$ counting the energy of each local magnetization under the demagnetizing field $\Hdm$ created by the surrounding magnetization, and the the Zeeman term $\fed{zm}$ counting the energy of each local magnetization to an applied external field $\Hx$. $\mu_0$ is the vacuum permeability and $\Ms$ is the saturation magnetization. For uniaxial magneto-crystalline anisotropy, $\Ku$ is a coefficient tuning the energetic barrier between two minima at $\mv\|\eau$ and at $\mv\|-\eau$, which can be tilted according to existing $\Hdm$ and $\Hx$. 

%Evolution of the magnetization configuration $\mv(\pos)$ under a cycling $\Hx$ was generally described by the Landau-Lifshitz-Gilbert equation, which is mathematically formulated as 
The governing equation for the evolution of local magnetization configuration $\mv(\pos)$ is formulated as
\begin{equation}
\begin{split}
&\dot{\mv}=\mv \times\left(\mv \times \varid{\fedf}{\mv} \right), \\
&\text{subject to}\quad|\mv|=1\label{eq:gov},
\end{split}
\end{equation}
which is also known as the Landau-Lifshitz equation with infinite sampling \cite{schabes1988magnetization, exl2014labonte, Miltat2007a}. Here $\dot{(\cdot)}=\partial(\cdot)/\partial t$.

% , yielding a quasi-equilibrium $\mv(\pos)$ at every incremental $\Hx$. 

Furthermore, since the exchange coefficient $\Ae$ is known to be tightly related to the domain wall energy at the equilibrium, the magnetic exchange coupling among grains and between grains as well as the GB-phase is worth noting, as it is dependent on the variation in the local structure and chemical composition. Such variation would also lead to energetic fluctuation within the microstructure during magnetization/demagnetization processes, manifested as the nucleation of a reversed magnetic domain or the impeded migration of the domain wall across the aforementioned interfaces. Such exchange coupling can be quantified by an effective field $\Hv^\mathrm{intf}_\mathrm{ex}$ as\cite{Vansteenkiste2014} 
\begin{equation}
    \Hv^\mathrm{intf}_\mathrm{ex} =\frac{2S}{\mu_0}\langle\frac{\Ae}{\Ms}\rangle\nabla\cdot\nabla\mv,
\end{equation}
where $\langle\cdot\rangle$ represents the harmonic mean of the quantity across an interface, and $S$ is a strength factor. When a complete exchange coupling is formed at the interface, $S=1$,  also implying a coherent atomic contact \cite{hernando1992role}. Although $S$ forms a well-defined parameter for implementing varying exchange coupling on a polycrystalline microstructure, in this work we assume $S=1$ for all existing interfaces. Instead, properties like $\Ae$, $\Ku$ and $\Ms$ are specified uniquely at the interface as it has been extracted as a unique region/phase (elaborated in \secref{sec:wf}). 

After obtaining the hysteresis, The (power) loss can be calculated as
\begin{equation}
    \PH=\omega\vol \WH\quad\text{with}\quad \WH=\oint H\diff B_\|,
    \label{eq:P_H}
\end{equation}
where $\omega$ stands for the frequency of the operating AC, $V$ is the volume of the core, and $\WH$ is an energy density characterizing the dissipated energy per unit volume within a complete loop. It is evident that each hysteresis loop corresponds to a unique $\WH$, thus $\WH$ can be used as a reference value for evaluating the hysteresis loss of a particular microstructure. $B_\|$ is the longitudinal magnetic induction (i.e., the projection of $\Bv$ along the direction of applied $\Hx$). As the  hysteresis results in merely the configurational evolution of the magnetization, (i.e., the $m$-$H$ loop), the following constitutive relation should be also considered
\begin{equation}
    \Bv=\mu_0\left[\Ms \mv + \Hv\right].
\end{equation}
Simulations of magnetic hysteresis can show variations, to reduce the influence of this on the calculated hysteresis loss, each hysteresis simulation contains 5-10 demagnetizing-magnetizing cycles. The magnetization was firstly reversed from
the positive saturation state to the negative one, then again reversed from the negative saturation state back to the positive one. Then, the hysteresis loss is calculated on the averaged $\BH$ loop
for each microstructure.

\subsection{Eddy current loss of a homogenized magnto-quasi-static system}\label{sec:ecl}

Eddy current loss in a magnetic core is the energy dissipation due to the electric impedance of the magnetic core when eddy current (induced by alternating magnetic field) flows through \cite{phillips1962classical}. Therefore, eddy current loss generally leads to reduced efficiency in the energy conversion of an electromagnetic device.

Based on the magneto-quasi-static (MQS) Maxwell's equation, the electrostatic and magnetic (vectorial) potentials should satisfy the following equations in a subdomain (with the volume $V$) inside a magnetic core
\cite{Albanese1990IEE} 
\begin{align} &\nabla\cdot\left(\econd\cdot{\Ev}\right)=0 \quad&\text{in }V, \\
    &\hat{\mathbf{n}}\cdot\llbracket\econd\cdot{\Ev}\rrbracket=0 \quad&\text{on }\partial V_\mathbf{intf}\cup \partial V,
    \label{eq:gov_ecl}
\end{align}
with an electrostatic field
\begin{equation*}
\begin{split}
    &{\Ev}=-\left(\nabla\phi + \dot{\mvp}\right)\nonumber, 
\end{split}    
\end{equation*}
where $\mvp$ is a vector potential of the exciting magnetic field, i.e., $\Bv \equiv \nabla\times\mvp$. 

\begin{figure}
\includegraphics[width=8.5cm]{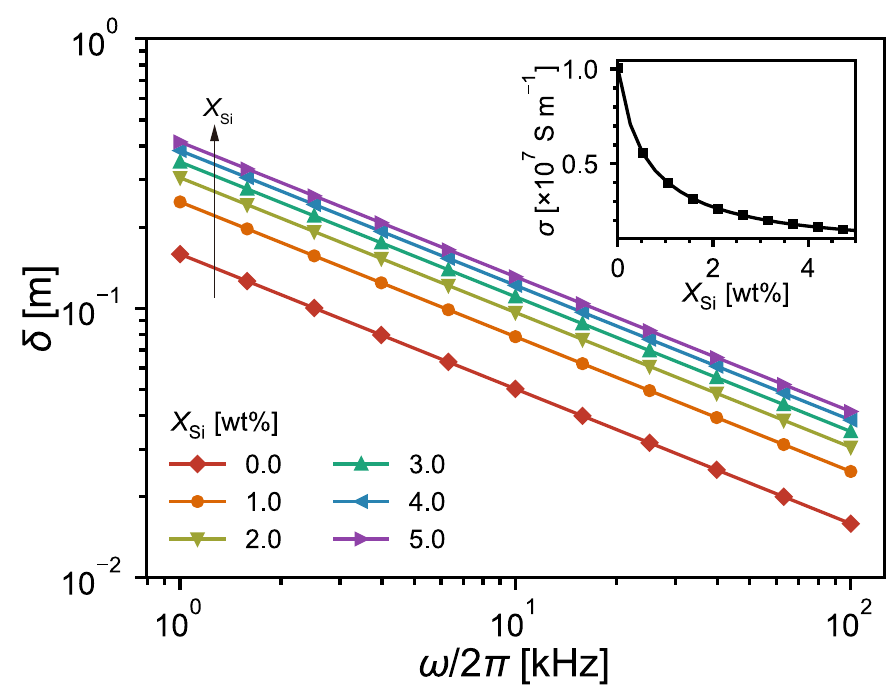}% Here is how to import EPS art
\caption{Skin depth $\delta$ of the $\fesi$ vs. operation frequency at various Si compositions $\XSi$. Inset: the $\XSi$-dependent electric conductivity $\sigma$ of the $\fesi$ \cite{numakura1972magnetic}.}
\label{fig:delta}
\end{figure}

When this alternating field is applied, the eddy current flows mainly around the surface of the magnetic core, characterized by a skin depth $\delta=\sqrt{2/(\mu \sigma\omega)}$. By taking the $\XSi$-dependent electric conductivity $\sigma$ of the $\fesi$ steel from Ref. \cite{numakura1972magnetic} we calculated this $\delta$ at various $\XSi$ in \figref{fig:delta} within a typical range of the excitation fields, i.e., $\omega/2\pi\in[0,100]~\si{kHz}$. As presented, the typical $\delta$ for $\fesi$ steel is on the magnitude of $10^{-3}\sim10^{-2}$ \si{m}. At higher frequencies, the skin effect becomes significant (i.e., decreasing $\delta$), yet it is still several order of magnitude larger than the typical length scale that resolves the mesoscopic microstructure (up to $10^{-4}$ \si{\meter}). Meanwhile, $\delta$ also increases with the rise of the Si compositions, mainly due to the drop of $\sigma(\XSi)$. This leads the eddy current problem in the mesoscopic cubic subdomain to a stationary conduction problem, and its (power) loss can be calculated by
\begin{equation}
    \begin{split}
        &P_\mathrm{E}=\int_{V} {\Ev}\cdot\econd\cdot{\Ev} \diff V= \dot{\mvp}\cdot\econd^\eff\cdot\dot{\mvp}V\label{eq:PE},
    \end{split}
\end{equation}
with an effective conductivity $\econd^\eff$. A detailed derivation has been provided in Ref. \cite{perna2020microstructure}. In this regard, knowing the spatial-temporal distribution of $\mvp$ can then lead us to the eddy current loss in the magnet. It is also worth noting that the mesoscopic microstructures affect this loss via $\econd^\eff$, reducing the eddy current problem of an AM-produced magnetic core to the homogenization of its electric conductivity under the assumption of homogeneous $\mvp$ in space. 

To find $\mvp$, we took the assumption that a mesoscopic cubic subdomain (edge lengths $L_x$, $L_y$, $L_z$ and volume $V=L_x L_y L_z$) is excited by a magnetic field $\mathbf{B}$ that is homogeneous in space while sinusoidal in time. An eddy current problem in this subdomain is defined by \cite{perna2020microstructure, Albanese1990IEE}
\begin{equation}
        \nabla\times\mathbf{B}=\mu_0{\Jv^\eff},\label{eq:ampere}
\end{equation}
where
\begin{equation*}
\begin{split}
    {\Jv}^\eff &= -\frac{1}{V}\int_V \econd\cdot{\Ev} \diff V = -\econd^\eff\cdot\dot{\mvp},\\
    \Bv&\equiv\nabla\times\mvp.
\end{split}
\end{equation*}
Meanwhile, for further simplification, the microstructural periodicity along the depth direction ($z$-direction) was tentatively neglected due to a limited dimension, i.e., $A_z = 0$. In this regard, adopting the analytical solution of $\mvp[\hat{\mathbf{r}}(x,y,0),t]$, \eqref{eq:PE} can be derived and formulated in a similar fashion as \eqref{eq:P_H}, i.e.,
\begin{equation}
    \begin{split}
        &P_\mathrm{E}=\omega V W_\mathrm{E}(\omega), \label{eq:pe}
    \end{split}
\end{equation}
where $W_\mathrm{E}$ is another energy density, formulated as
\begin{equation}
    W_\mathrm{E}(\omega)=\frac{4B^2_0}{\mu_0}\sum_r^{x,y}\sum_k^\infty \frac{\Pi_{rk}}{\gamma_k^2 \sqrt{\left(\frac{\gamma_k \delta_r}{\sqrt{2} L_r}\right)^4 +1 }}\label{eq:we}
\end{equation}
with
\begin{equation*}
    \begin{split}
        \gamma_{k}&=(2k-1)\pi,\\
        \Pi_{xk} &= \frac{\lambda^\Im_{xk}\sinh\lambda^\Re_{xk}L_{y}-{\lambda^\Re_{xk}\sin\lambda^\Im_{xk}L_{y}}}{\lambda^\Re_{xk}\lambda^\Im_{xk}L_{y}(\cosh\lambda^\Re_{xk}L_{y}+\cos\lambda^\Im_{xk}L_{y})},\\
        \lambda_{xk}&\equiv \lambda^\Re_{xk} + \lambda^\Im_{xk} \iu=\frac{\sqrt{2}}{\delta_x}\sqrt{\left(\frac{\gamma_k\delta_x}{\sqrt{2}L_x}\right)^2+\iu},\\
        \delta_x&=\sqrt{\frac{2}{\mu_0\omega\sigma_{xx}^\eff}}.
    \end{split}
\end{equation*}
Here the component index $r=x,~y$, and $\iu$ is the imaginary unit.
The expressions of $\Pi_{yk}$,  $\lambda_{yk}$ and $\delta_y$ are obtained by exchanging the subscripts $x\leftrightarrow y$. The skin depth along $r$-direction $\delta_r$ is calculated by the component $\sigma_{rr}$ of the conductivity tensor $\econd$. Unlike $W_\mathrm{H}$ for the hysteresis loss, $W_\mathrm{E}$ explicitly presents the dependence of $\omega$ due to the frequency-dependent nature of the skin effect. 
As for the chosen subdomain is sufficiently smaller than the skin depth of any directions, i.e., $L_r<<\delta_r$, the low frequency approximation can be adopted for \eqref{eq:we}, yielding
\begin{equation}
    W_\mathrm{E}\approx W_\mathrm{E}^\text{lf}=\frac{4B_0^2 \omega}{ L_x L_y}\sum^{x,y}_r \sigma^\eff_{rr}L^4_r \Pi_r,
    \label{eq:we_lf}
\end{equation}
with
\begin{equation*}
    \Pi_x=\frac{\sinh\left(\pi\frac{L_{y}}{L_{x}}\right)-\pi\frac{L_{y}}{L_x}}{\cosh\left(\pi\frac{L_{y}}{L_x}\right)-1}.
\end{equation*}
It should be noticed that $W_\mathrm{E}^\text{lf}$ has the dimension of energy per volume, characterizing the eddy current loss inside the material.

Based on \eqref{eq:we_lf} the eddy current problem within a mesoscopic cubic subdomain can be degenerated to a stationary conduction problem for $\fesi$ steel with the operation frequency up to 100 Hz, when the assumption of a spatially-homogeneous $\mvp$ can be safely adopted. Considering the following decomposition for the potential $\phi$
\begin{equation}
    \phi=\sum^3_i a_i\phi_i,\quad\mathrm{where}\quad\dot{\mvp}\equiv\sum_i^3 a_i\cuv
    \label{eq:decomp}
\end{equation}
with $\cuv$ the Cartesian unit vector of the subdomain. Then, the governing equation shown in \eqref{eq:gov_ecl} can be re-written as
\begin{align}
&\nabla\cdot\left(\econd\cdot\hat{\Ev}_i\right)=0 \quad&\text{in }V, \\
&\hat{\mathbf{n}}\cdot\llbracket\econd\cdot\hat{\Ev}_i\rrbracket=0 \quad&\text{on }\partial V_\mathbf{intf}\cup \partial V,\\
    &\text{with}~~\hat{\Ev}_i=-\left(\nabla\phi_i + \cuv\right)\nonumber,
    \label{eq:gov_ecl}
\end{align}
In this regard, we define the following homogenization problem on each direction (labelled by index $i=x,y$)
\begin{equation}
    \left\{\begin{aligned}
    \Jv_i & =-\econd\cdot\hat{\Ev}_i & & \text { on microscale } \\
    \left\langle\Jv_i\right\rangle & =-\econd^\mathrm{eff}\cdot\left\langle \hat{\Ev}_i\right\rangle & & \text { on macroscale }
    \end{aligned}\right.
    \label{eq:homogen}
\end{equation}
with the Hill condition
\begin{equation}
    -\left\langle\Jv_i\cdot\hat{\Ev}_i\right\rangle=-\left\langle\Jv_i\right\rangle\cdot\left\langle\hat{\Ev}_i\right\rangle=\left\langle\Jv_i\right\rangle\cdot\left\langle-\hat{\Ev}_i\right\rangle.
    \label{eq:hill}
\end{equation}
Here $\langle\cdot\rangle=\int_V(\cdot)\diff V / V$. Taking the scalars $\bar{J}^i_j$, and $\bar{E}^i_j$ ($j=x,y$) as the components of the vector $\left\langle\Jv_i\right\rangle$ and $\left\langle \hat{\Ev}_i\right\rangle$, the components of $\econd^\eff$ are then calculated as
\begin{equation}
    \sigma^\eff_{ij}={\bar{J}^i_i}/{\bar{E}^j_j}.
\end{equation}

As its tensorial form suggested, the eddy current loss of the AM-produced magnetic core can be affected by both the local ($\econd$) and the overall anisotropies ($\econd^\eff$) of the mesoscopic microstructure, besides the influences from Si compositions and operation frequency. Notably, it is expected that the overall anisotropy in conductivity of the AM-produced magnetic core is significantly affected by polycrystalline texture as well as the formation of GB-phase. 
%\denotedtext{some discussion of how the polycrystalline texture of AM-produced sample will be anisotropic, e.g., columnar grain}.
The behavior of soft magnetic materials is significantly influenced by microstructure and texture, i.e. high permeability correlates with a pronounced <001> direction. For the microstructure, no preferential grain orientation was reported, as the sintering in the BJP-process does not change orientations. Meanwhile, due to crystallographic incoherence, enhanced scattering of the carriers (like electrons) can also be expected to cross GB phases, resulting in lowered local conductivity. Such reduced conductivity (increased resistivity) in GB-phases can further enhance the anisotropy originating from polycrystalline texture, as the eddy current would be limited within the grain interior region. A similar phenomenon has been unveiled for the homogenized thermal conductivity of a mesoscopic microstructure with the thermal resistance of the interface considered \cite{yang2022diffuse}.

\section{Simulation setup \& parameters}
\subsection{Demagnetization simulation and magnetic properties}

The Landau-Lifshitz equation in \eqref{eq:gov} has been implemented in the package $\mumax$ by FDM \cite{Vansteenkiste2014, exl2014labonte}. It has been shown that \eqref{eq:gov} is in accordance with the iterative scheme of steepest conjugate gradient (SCG) method \cite{exl2014labonte, furuya2015semi} for constrained optimization of $\fedf$, which can easily be implemented numerically. The simulation domains of all synthetic microstructures and the digitized sample-B microstructure have a size of $512\times512\times2~\si{\micro\m^3}$, while the simulation domain of SEM-digitized microstructure of sample A is $240\times240\times2~\si{\micro\m^3}$ due to the limited size of the image. The FD grid size is $1\times1\times1~\si{\micro\m^3}$. Periodic boundary conditions (PBC) were applied along
the out-of-plane (z) direction by macro geometry approach \cite{fangohr2009new}, while Neumann boundary conditions were applied on
other boundaries \cite{Vansteenkiste2014}. It should be noted that the simulation domain is equivalent to an intersection of a long elliptic
cylindrical structure with columnar nanocrystals, where the in-plane domain configuration and domain wall migration
are mainly resolved. In that sense, the grid number along the out-of-plane direction was decreased for the reduction of computational consumption.

\begin{figure}
\includegraphics[width=8.5cm]{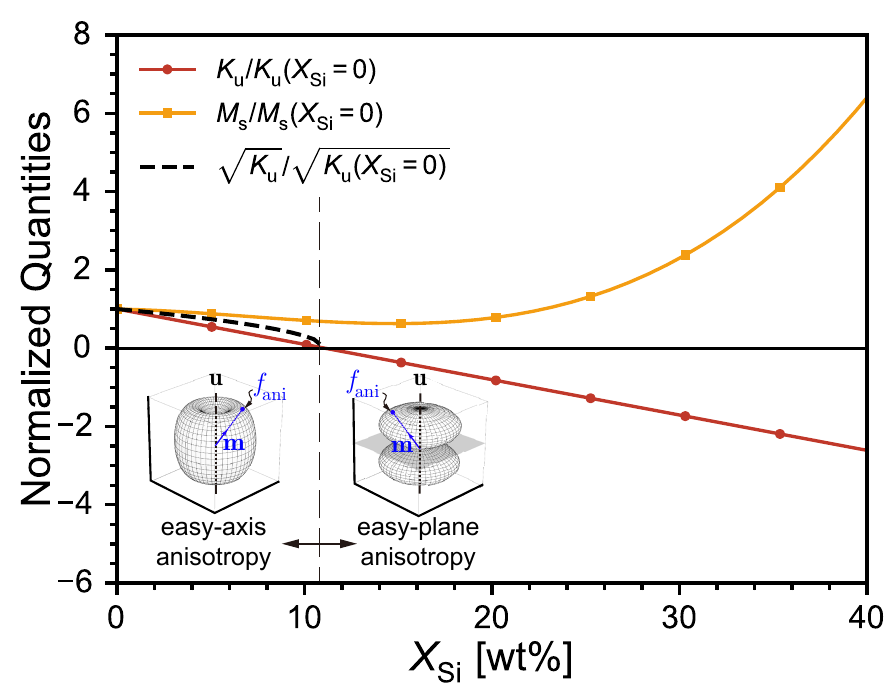}% Here is how to import EPS art
\caption{Concentration dependence of the magnetocrystalline anisotropy and saturation magnetization.}
\label{fig:mmparams}
\end{figure}

Magnetic properties of electrical steel present a dependence of its chemical composition, specifically, $\Ku$ and $\Ms$ can be formulated as functions of the \ce{Si} fraction (in wt\%)\cite{antonio2015some}, i.e., 
\begin{equation}
    \begin{split}
        \Ku =& 48-433.3 \XSi~[\si{kJ~m^{-3}}], \\
        \Ms  =& (1.717-1.639\XSi-65.253\XSi^2+318.308\XSi^3)\E{3}~[\si{kA~m^{-1}}].
    \end{split}
    \label{eq:ku-ms}
\end{equation}
Meanwhile, $\Ku$ and $\Ms$ in the GB-phase are scaled down with a factor of $0.001$. This modification is done in order to approximate the unknown properties of the lamellar FeSi/Fe$_2$B mixed region.
Structural and magnetic parameters of the microstructures observed from samples A and B are listed in \tabref{tab:params}. In \figref{fig:mmparams} the Si-dependence of $\Ku$, $\Ms$ and $\sqrt{\Ku}$ are presented. It is worth noting that the magnetocrystalline anisotropy of the material switches from easy-axis type to easy-plane one when $\XSi\geq~11\%$, as $\Ku<0$. At that moment the spontaneous magnetization prefers the in-plane directions that are perpendicular to the former easy-axis $\mathbf{u}$. In addition, as the Bloch domain wall energy $\sigb\propto\sqrt{\Ku}$ and the Néel domain wall energy $\sigb\propto\Ms$ when considering a Si-independent exchange stiffness $\Ae$ \cite{kronmuller2003micromagnetism}, \figref{fig:mmparams} also depicts the tendency of domain wall energy w.r.t. $\XSi$, implying energy fluctuations during domain wall motion across GB region where distinct $\XSi$ is expected. Taking the assumed $\Ku^\mathrm{GB}=0.001\Ku^\mathrm{G}$ as an instance, when reversed domain wall readily nucleated at GB region and about to move into a grain, extra energy, provided by inversely increased external field, is expected. This domain wall is then ``pinned'' at the GB-phase and thus influences the coercivity and eventual hysteresis loss of the material. Materials and microstructure parameters of sample A and B are correspondingly listed in \tabref{tab:params}.

Scale bridging between the domain wall motion and the observed microstructure is another challenge in conducting mesoscopic polycrystalline demagnetization simulation with a physically-sound interpretation. The typical length scale of the domain wall dynamics, characterized by the Bloch domain wall thickness $\ldw$, in the range of tens nanometers for electrical steel. Meanwhile, to guarantee the numerical validity of Landau-Lifshitz dynamics by FDM, sufficient grid amount (normally $\geq 4$) should be reserved across the domain wall, where the magnetization vectors vary in a diffuse-interface fashion \cite{kronmuller2003micromagnetism}. In this regard, a proper demagnetization simulation on a mesoscopic microstructure with physically-sound parameters would impose gigantic requirements on computational power and efficiency. As a compromise, to firstly guarantee the numerical validity, we assign a Bloch domain wall thickness $\ldw\sim4~\si{\micro m}$ at zero $\XSi$, which is readily in the range of permalloy \cite{coey2010magnetism}. According to the correlation between $\ldw$ and Landau-Liftshitz model parameters, i.e., 
\begin{equation}
    \ldw=\pi\sqrt{\frac{\Ae}{\Ku}}.\label{eq:ldw}
\end{equation}
one can inversely calculate a gradient constant $\Ae$ that is about 5,000 times larger than the physical exchange coefficient ($A_\mathrm{e}$) of the realistic electrical steel \cite{edstrom2015magnetic, kuz2020exchange}. In that sense, we have to note the employed material in the simulation should be identified as a model material, which embodies the $\XSi$-dependencies of $\Ku$ and $\Ms$ as electrical steel while presenting similar domain wall level behaviors as permalloy. Nonetheless, the constructed model material should be sufficient to deliver valuable insights regarding interactions between domain wall dynamics (nucleation, migration, pinning, etc.) and the microstructural characteristics (grain size, grain boundaries, GB-phase, etc), and to reproduce tendencies of magnetic properties ($\Hc$, $\Br$, and hysteresis loss $\WH$) w.r.t. the varying microstructural descriptors and compositions.

\begin{table}[!h]\centering
\caption{Materials and microstructure parameters.}
\small
\begin{tabular}{cccc}
\hline
Parameter & Sample A & Sample B& Unit  \\ \hline
$\XSi$ & 3 & 5 & [wt\%] \\  
$\gl$ & 48 & 149 & [$\si{\micro\meter}$]\\
$\gbw$ & 1 & 12 & [$\si{\micro\meter}$]\\
$\sigma^\mathrm{G}$ & $2.076\E{6}$ &  $1.477\E{6}$ & [$\si{S~m^{-1}}$]\\
$\sigma^\mathrm{GB}$ & $0.018$ &   $0.018$ & [$\si{S~m^{-1}}$]\\
$\Ku^\text{G}$ & 35.001 & 26.335 & [$\si{kJ~m^{-3}}$]\\
$\Ku^\text{GB}$& 0.035 & 0.026 &[$\si{kJ~m^{-3}}$]\\
$\Ms^\text{G}$& 1617.960 &1511.96 & [$\si{kA~m^{-1}}$]\\
$\Ms^\text{GB}$& 1.618 & 1.512 &[$\si{kA~m^{-1}}$]\\\hline
\end{tabular}
\label{tab:params}
\end{table}

Demagnetization simulations were also performed on the microstructures with continuously changing $\gl$ and $\gbw$ to investigate the microstructural dependencies of magnetic properties ($\Hc$, $\Br$ and $\WH$). In this case, 
synthetic microstructures were generated and grouped into two batches. The first batch ($\gl$-batch) contains a series of microstructures with gridded $\gl$, ranging from 60 to 160 \si{\micro m}, and a fixed $\gbw=5$ \si{\micro m}. The second batch ($\gbw$-batch) contains a series of microstructures with gridded $\gbw$, ranging from 1 to 20 \si{\micro m}, and a fixed $\gl=149$ \si{\micro m}. Parameters of two distinct $\XSi$ values (as listed in \tabref{tab:params}) were also separately adopted and examined.

\subsection{Computational homogenization of stationary conductivity}

The deduced stationary conduction problem in \eqref{eq:gov_ecl} has been implemented by FEM within the program \texttt{MFM-FM}, developed by the authors based on the MOOSE framework (Idaho National Laboratory, ID, USA) \cite{permann2020moose}. 
The FE simulation domains have the identical construction as the FD grids used in the hysteresis simulation (\secref{sec:fd_setup}), while the 8-node hexahedron Lagrangian elements were employed. This also means that simulation domains for homogenization of all synthetic microstructures and the digitized sample-B microstructure have a size of $512\times512\times2~\si{\micro\m^3}$, while the simulation domain of SEM-digitized microstructure of sample A is $240\times240\times2~\si{\micro\m^3}$ due to the limited size of the image. A nonlinear
steady solver with the preconditioned Jacobian-free Newton-Krylov method (PJFNK) has been employed. As both the digitized and reconstructed microstructures in this work are not strictly periodic, the direct homogenization scheme was carried out. The linear potential BCs are practically employed on $\phi_i$ ($i=x,y$) to guarantee the Hills condition in \eqref{eq:hill}, reading as
\begin{align} 
&\phi_i=\hat{\Ev}_i\cdot\hat{\mathbf{r}},
 \quad&\text{on }\partial V,
\end{align}
where $\hat{\mathbf{r}}$ is the coordinate, $\hat{\mathbf{r}}\in V$, and $\hat{\Ev}_i$ is the prescribed constant vector. As $\cuv$ are constant over the whole simulation domain, it is then easy to prove that 
\begin{equation*}
    \langle\nabla\phi_i\rangle=\frac{1}{V}\int_{\partial V}\phi_i\hat{\mathbf{n}}\diff S=
    \frac{\hat{\Ev}_i}{V}\int_{\partial V}\hat{\mathbf{r}}\cdot\hat{\mathbf{n}} \diff S=\hat{\Ev}_i,
\end{equation*}
with the norm vector $\hat{\mathbf{n}}$ on $\partial V$.

The $\XSi$-dependent conductivity for the grain interior is taking from Ref. \cite{numakura1972magnetic} and fitted by inversed \nth{2}-order polynomial as
\begin{equation}
\begin{split}
        \econd^\mathrm{G} = &\mathbf{I}(-562.8\XSi^2 +142.6\XSi + 1.0)^{-1}\E{7}~[\si{S~m^{-1}}],\\
\end{split}
\end{equation}
where $\mathbf{I}$ is the rank-2 identity tensor. Meanwhile, as it is difficult to obtain the exact conductivity in the GB-phase due to limited dimensions, we assume the local conductivity in the GB-phase to be identical to a bulk \ce{Fe2B}, measured as 0.018 $\si{S~m^{-1}}$ \cite{zhao2020synthesis}.

Although simulation domains of two distinct sizes were employed in the FE homogenization of $\econd^\eff$, it should be noted that the identical subdomain size was used in calculating the eddy current loss based on \eqref{eq:pe} and \eqref{eq:we}, i.e., $L_x=L_y=512$ \si{nm} and $L_z=2$ \si{nm}.

\section{Result and discussion}
\subsection{Hysteresis features and hysteresis loss}

\begin{figure*}[!t]
	\centering
  \includegraphics[width=18cm]{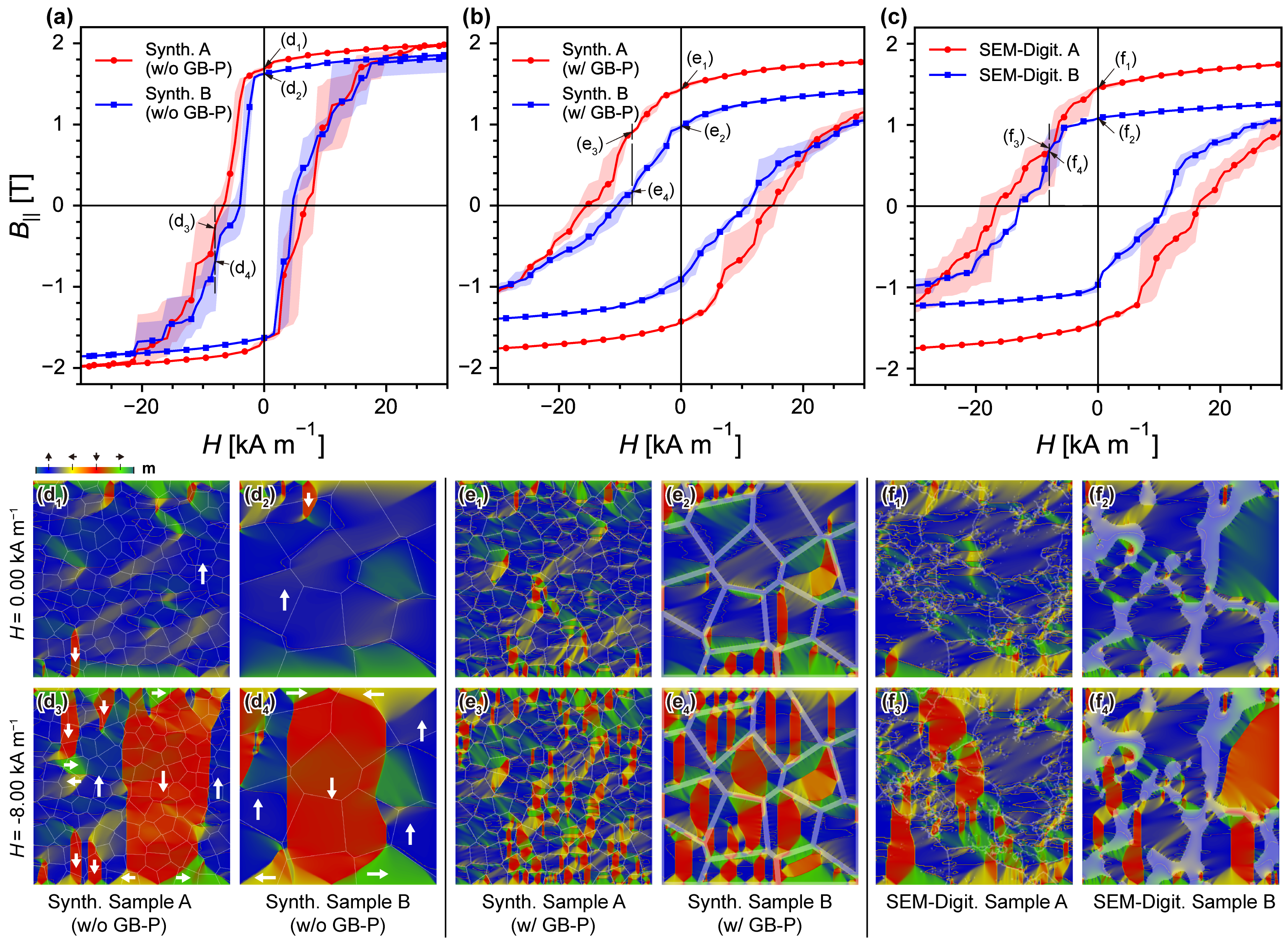}
	\caption{Simulated hysteresis: (a) synthetic microstructures without GB-phase, (b) synthetic microstructures with GB-phase, (c) SEM-Digitized microstructures. Domain structures are also illustrated on (\ce{d1})-(\ce{d4}) synthetic microstructures without GB-phase, (\ce{e1})-(\ce{e4}) synthetic microstructures with GB-phase, and (\ce{f1})-(\ce{f4}) SEM-digitized microstructures, at $H=0.0~\si{kA~m^{-1}}$ and $H=-8.0~\si{kA~m^{-1}}$, respectively. }
	\label{fig:hyst}
\end{figure*}

Fig. \ref{fig:hyst} shows the simulation setup and results of samples A and B, for different reconstructions. This includes reconstructions w/o GB-phase Fig. \ref{fig:hyst} (a), w/ GB-phase Fig. \ref{fig:hyst} (b) and SEM-digitized ones Fig. \ref{fig:hyst} (c). Shaded areas indicate the deviations in the simulations over multiple runs. 
Reconstructions w/o GB show similar hysteresis behavior, though differences can be found in the coercive fields and remanence. Simulated coercive fields for sample A and B could be determined as $6.8~$ and $4.3~\si{kAm^{-1}}$, roughly a difference of $40~\%$. Differences in the remanence are less pronounced with sample A showing $\Br$ as $1.7~\si{T}$ compared to sample B with $\Br$ $1.7~\si{T}$. Overall the hysteresis behavior can be considered hard with a sudden reversal close to the coercive field and only minor switching events before that. 

Introducing the grain boundary phase modifies the behavior of the hysteresis significantly, increasing  the coercive fields of sample A and B to $15.2$ and $10.7~\si{kAm^{-1}}$. Notably the ratio of simulated coercive fields $H_{cA}/H_{cB}=1.42$ is comparable to the ratio of the experiments \cite{kumari2023improving}, which is $H_{cA}/H_{cB}=1.49$. Both samples show a reduction in $\Br$ and $\Bs$, though more pronounced in sample B, as can be expected due to the reduced properties of the GB-phase and its larger volume in sample B. The shape of the hysteresis loop has changed to resemble a softer change for both cases,  with the reversal happening gradual compared to reconstructions w/o GB-phase. In the case of SEM-digitized samples A and B (see Fig. \ref{fig:hyst} (c)) similarities to the synthetic reconstructions with GB-phase can be found, this can be seen in the shape of the hysteresis loop and the simulated properties of the structures, such as $\Hc$ and $\Br$. With simulated results of coercive fields at $16.5~$ and $11.9~\si{kAm^{-1}}$ for sample A and B, respectively.

A more detailed view into the magnetic behavior of the material is provided in Fig. \ref{fig:hyst} (d$_1$ -f$_4$), a visualization of the magnetic domains at external fields of 0$~\si{kA/\meter}$ (images on the top) and -8$~\si{kA/\meter}$, the remanent state (images on the bottom). In the case w/o GB-phase \ref{fig:hyst} (d$_1$-d$_4$) very little change can be seen at 0 $~\si{kA/\meter}$, small domains have already emerged, though no preferred nucleation sites could be identified. It should be noted, that most of the complete and partial rotation is focused on the top and bottom of the simulation domain. At -8 $~\si{kA/\meter}$ the existing domains have expanded significantly and new ones emerged in the system. The observed domains do not follow the grain boundaries, but instead cover multiple grains each, with completely switched domains neighboring multiple partially switched domains, especially towards their top and bottom. Introducing grain boundaries into the system changes the behavior significantly, see Fig. \ref{fig:hyst} (e$_1$-e$_4$), at 0 $~\si{kA/\meter}$ more domains have emerged compared to  Fig. \ref{fig:hyst} (d$_1$,d$_2$) the completely switched domains are confined to single grains, whereas partially switched ones cross the grain boundaries regularly. A similar behavior can be observed in Fig. \ref{fig:hyst} (e$_3$), where a large number of switched domains is distributed throughout the system, mostly confined to single grains, though some domains cross the grain boundary. Fig. \ref{fig:hyst} (e$_4$), shows comparable behavior though the singular domains are larger, due to an increased $\gl$. Some cases can be seen, where switched domains either crossed the boundaries or connected to domains inside neighboring grains. An overall trend can be observed, where domains expand parallel to $H$. Lastly SEM-digitized samples were simulated as can be seen in Fig. \ref{fig:hyst}(f$_1$-f$_4$) (f$_1$) shows isolated switched regions, yet no direct correlation to local features is visible. In contrast (f$_2$) shows switched domains preferably in areas fully enclosed by GB-phases. Under an applied electric field different behaviors can be observed. Sample A shows expansion of domains and the nucleation of additional ones, in this case the domains can stretch over multiple grains. Sample B shown in (f$_4$) on the other hand exhibits domains mainly restricted to singular grains. The smaller domains, surrounded by GB-phases remain mostly unchanged.

\begin{figure*}[!t]
	\centering
 	\includegraphics[width=18cm]{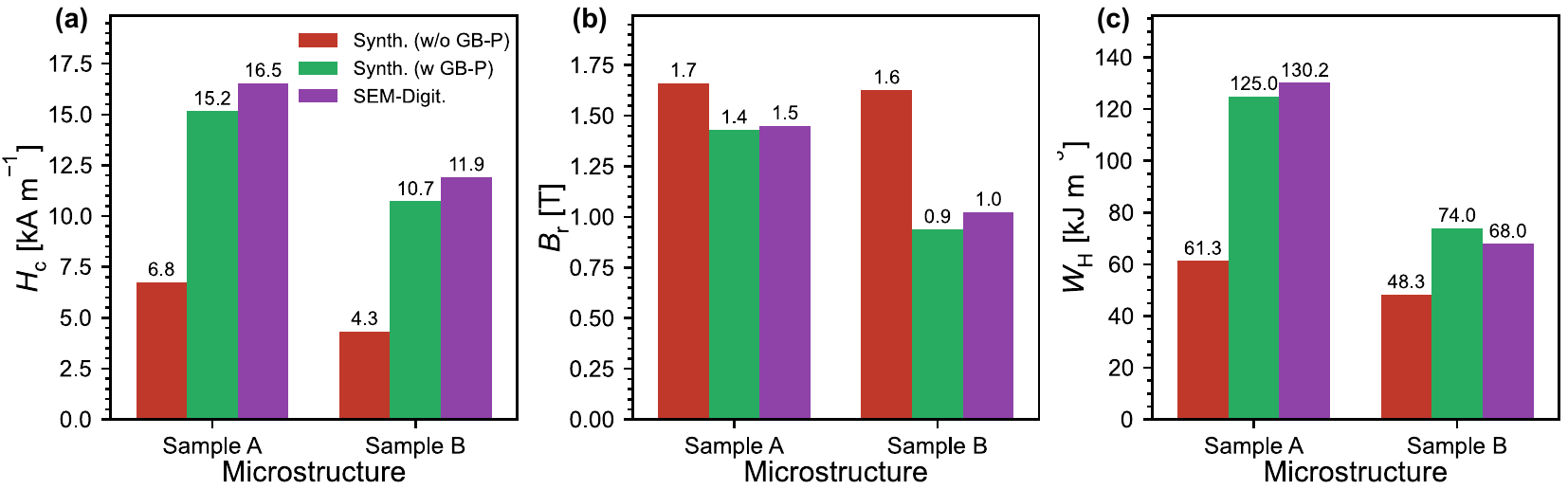}
	\caption{Hysteresis features, including (a) coercivity $\Hc$, (b) remanence $\Br$ and (c) hysteresis loss $\WH$. Red indicates a synthetic microstructure w/o GB-phase, green a synthetic microstructure w GB-phase and purple direct digitization from SEM-photomicrographs.}
	\label{fig:hyst_loss}
\end{figure*}

\begin{figure*}[!t]
	\centering
 	\includegraphics[width=18cm]{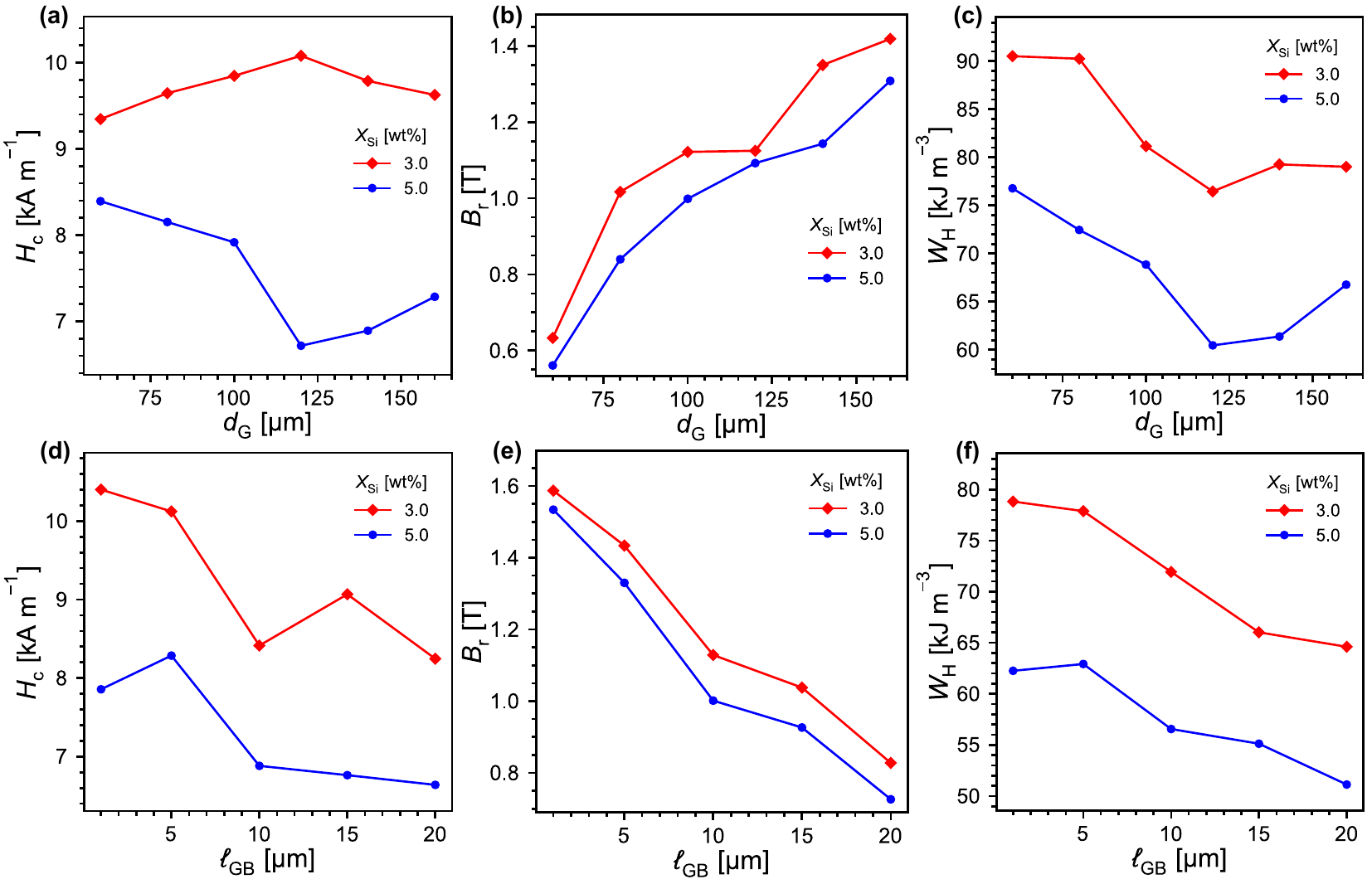}
	\caption{Hysteresis features vs. microstructure descriptors: (a) coercivity $\Hc$ (b) remanence $\Br$ and (c) hysteresis loss $\WH$ vs. grain size $\gl$, and (d) $\Hc$, (e) $\Br$ and (f) $\WH$ vs. GB-phase thickness $\gbw$.}
	\label{fig:hyst_PS}
\end{figure*}

The different resulting magnetic properties, in particular $\Hc$, $\Br$ and $\WH$ are visualized in Fig. \ref{fig:hyst_loss}, red, green and purple bars represent the results for a synthetic reconstruction w/o GB-phase, with GB-phase and SEM-digitized respectively. It can be seen that the introduction of the GB-phase changes the magnetic properties significantly. $\Hc$ more than doubles with the introduction of the GB-phase for sample A and B, though it should be noted that SEM-digitized samples show $\Hc$ slightly larger compared to the synthetic reconstructions w /GB. $\Br$ also showed a response to the introduction of GB-phase, as can be seen in fig \ref{fig:hyst_loss} (b), sample A w/o GB-phase showed the largest remanence, introducing the GB-phase reduces this slightly from $1.7$ to $1.4~\si{T}$, though they are still of comparable magnitude. In sample B the reduction of remanence was more significant, in this case the GB-phase almost halves $\Br$ from $1.6$ to $0.9~\si{T}$. This can be attributed to the larger GB-phase used in sample B, as effective properties of GB-phase are assumed to be small in comparison to the pure material. Results for SEM-Digitized microstructures are close to results obtained from synthetic samples w/ GB. Fig. \ref{fig:hyst_loss} (c) shows $\WH$ for the different reconstructions w/o GB-phase the losses are low with $61.3$ and $48.3~\si{kJ\meter^{-3}}$, the GB-phase doubles the losses for sample A and increases $\WH$ by ~$50~\%$for sample B.\newline
To further investigate the microstructure property relation several sets of simulations were performed, the extracted properties are visualized in Fig. \ref{fig:hyst_PS}, with (a-c) being the variation of $\gl$ with a constant $\gbw$. Fig. \ref{fig:hyst_PS} (d-f) shows the variation of $\gbw$ with constant $\gl$. The $\gl$ series shows two different coercivity-trends with samples of $\XSi=3.0$ wt\% and the ones with  $\XSi=5.0$ wt.\%, samples with lower $\XSi$ show a light trend of increasing coercivity, whereas the samples with $5.0$ wt\% show a decreasing $\Hc$ with increasing grain size. With regards to $\Br$ all simulations show a consistent trend of increasing $\Br$ with increasing $\gl$. This can be attributed to the change in the volume fraction of the GB-phase, as the grain size increases the fraction of GB-phase, which has a reduced $\Bs$, decreases. $\WH$ shows an overall negative trend with increasing $\gl$, though the overall lowest losses do not occur for the largest grain sizes, but instead they can be observed at $\gl=120~\si{\micro\meter}$.\newline
The variation of $\gbw$ plots another picture, while for $\XSi=3.0$ wt\% a rough trend of declining $\Hc$ can be observed, the same cannot be said for the other series, with too much variation to establish a clear trend. $\Br$ decreases with increasing $\gbw$, as it mostly depends on the volume fraction of the GB-phase. Hysteresis losses also show a trend of decreasing $\WH$ with increasing $\gbw$. 

\subsection{Stationary conductivity and eddy current loss}

\begin{figure*}[!t]
	\centering
 	\includegraphics[width=18cm]{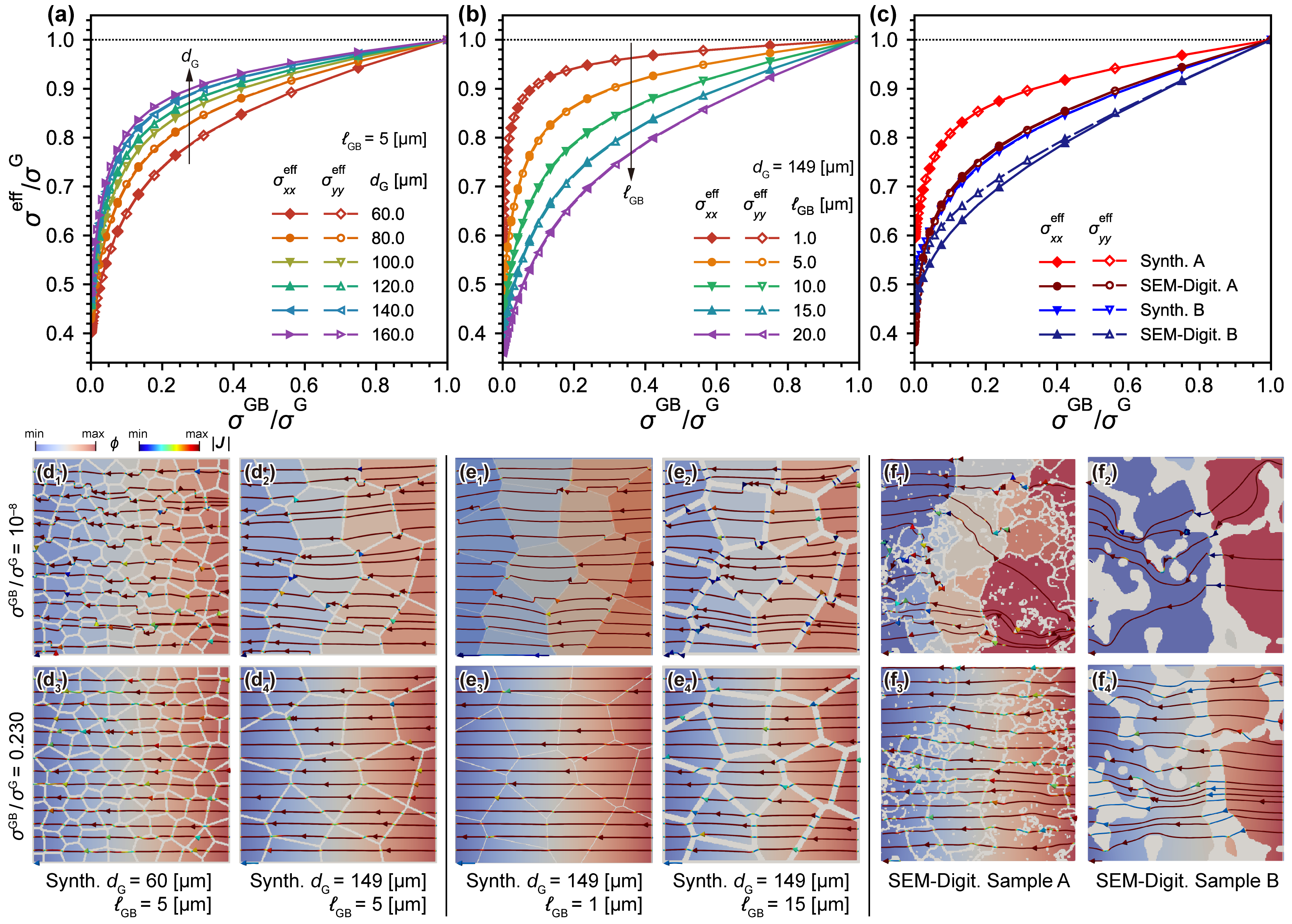}
	\caption{Diagonal components of the effective conductivity tensor of microstructures vs. varying conductivity value of the GB-phase. The effective conductivity is homogenized on (a) the synthetic microstructures with varying grain size $\gl$, (b)  the synthetic microstructures varying GB-phase thickness $\gbw$, and (c) synthetic / SEM-digitized microstructures of different samples. Potential contours along with the electric currents are also illustrated on (\ce{d1})-(\ce{d4}) synthetic microstructures with varying grain size $\gl$, (\ce{e1})-(\ce{e4}) synthetic microstructures with varying grain size $\gbw$, and (\ce{f1})-(\ce{f4}) SEM-digitized microstructures of different samples, for $\sigma^\mathrm{GB}/\sigma^\mathrm{G}=1\E{-8}$ and $\sigma^\mathrm{GB}/\sigma^\mathrm{G}=0.23$, respectively. }
	\label{fig:sigma_eff}
\end{figure*}

Fig. \ref{fig:sigma_eff} shows the normalized effective conductivities of varying synthetic samples. Wherein Fig. \ref{fig:sigma_eff} (a) depicts samples with a constant grain boundary thickness of 5$~\si{\micro\meter}$ and Fig. \ref{fig:sigma_eff} (b) shows the variation of grain boundary thickness for a constant $\gbw$ of 149$~\si{\micro \meter}$. Also analyzed were the SEM-digitized microstructures, their conductivites are shown in Fig. \ref{fig:sigma_eff} (c).
The effective conductivity is influenced by the spatial distribution of grains and corresponding boundary phases. This is represented in Fig. \ref{fig:sigma_eff} (a) as $\sigma^\eff$ shows larger values for larger grain sizes and approaches $\sigma^G$ faster, though it should be noted, that the differences between conductivity values decrease as the grain size increases. 

A variation of GB-thickness shows a trend of decreasing $\sigma^\eff$ with increasing $\gbw$. Different behaviors can be observed for different $\gbw$, namely, a thin GB results in a behavior of $\sigma^\eff$ nearly box shaped, but flattening out with an increase in $\gbw$. 

These different behaviors can be attributed to the changes in the overall setup of the system, where an expansion of the GB region translates to a reduction in conductivity. As grain size increases the relative volume fraction of GB regions is reduced. The inverse of this effect can be seen for the variation of the GB-thickness, wherein the fraction of GB regions increases with thickness. In both cases $\sigma^\eff$ was evaluated for different directions resulting in separate values for $\sigma_{xx}^\eff$ and $\sigma_{yy}^\eff$, though in this specific case the results are nearly identical. This can be attributed to the reconstruction route, as the microstructure is based on Poisson-disk sampling.

\begin{figure*}
\includegraphics[width=18cm]{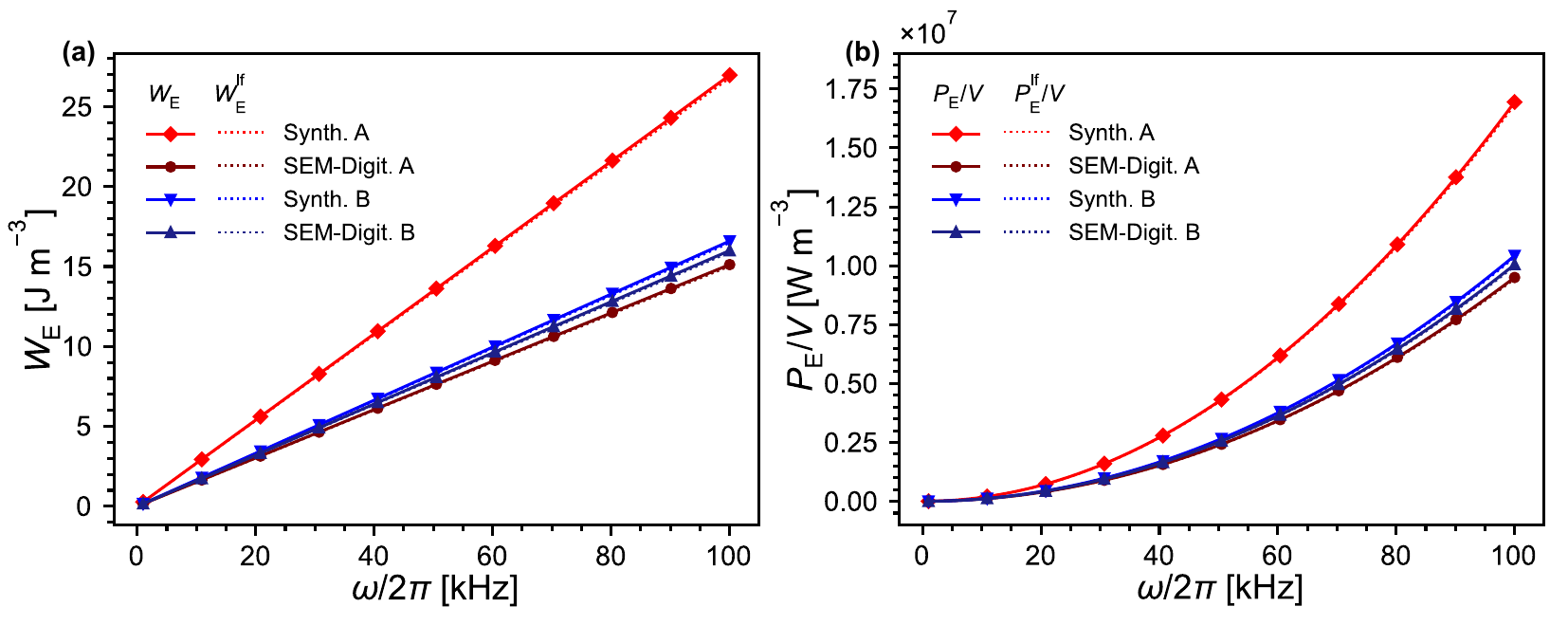}% Here is how to import EPS art
\caption{Eddy current losses in (a) energy density $\WE$, and (b) power density $P_\mathrm{E}/V$ by \eqref{eq:we} vs. operation frequency in a $512\times512\times2$ \si{\micro m^3} subdomain with effective conductivity, obtained by various reconstructed microstructures. The eddy current losses at low-frequency limit by \eqref{eq:we_lf} are also depicted. }
\label{fig:we}
\end{figure*}

\begin{figure*}
\includegraphics[width=18cm]{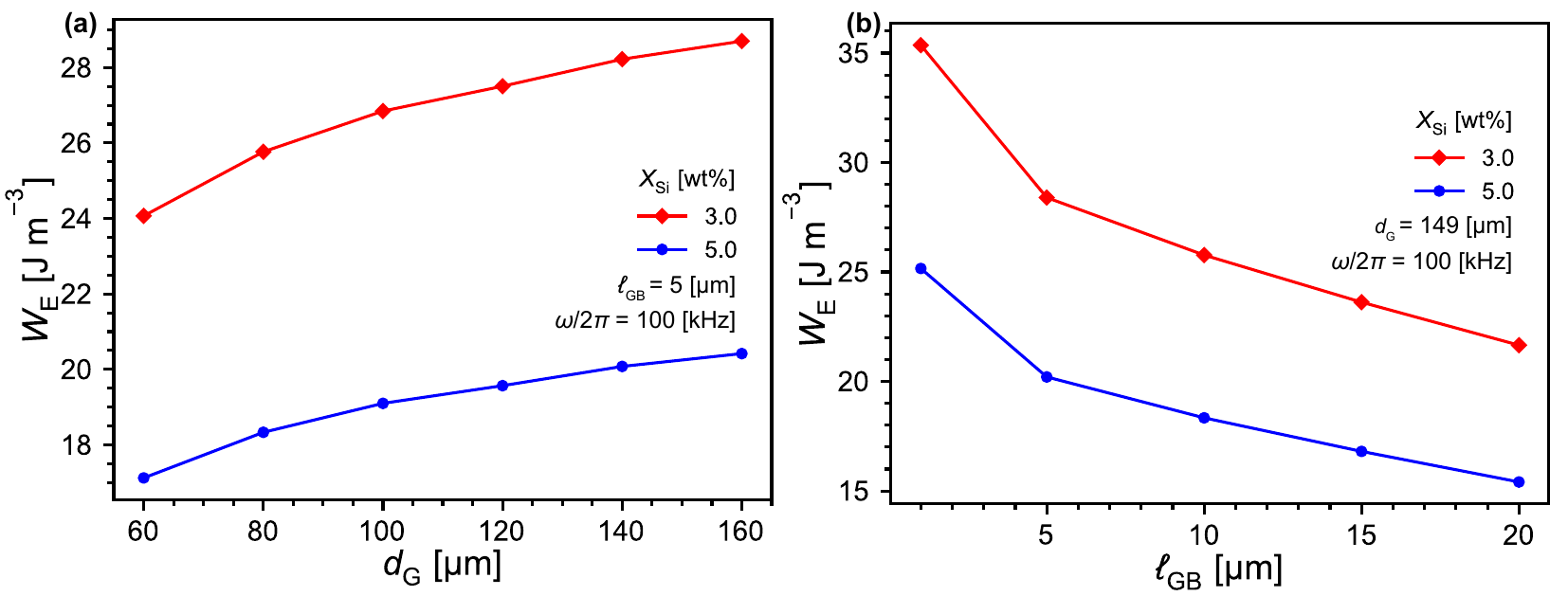}% Here is how to import EPS art
\caption{Eddy current losses at 100 kHz. vs. (a) varying grain size $\gl$, and (b) varying GB-phase thickness $\gbw$. }
\label{fig:we_var}
\end{figure*}

Fig. \ref{fig:sigma_eff} (d) - (f) gives a detailed view of the influence of the electric properties of the microstructure. In particular the potential $\phi$ and the electric currents $|J|$. The top row illustrates the case of a resistive GB-phase, with a ration of $\sigma^{\mathrm{GB}}/\sigma^{\mathrm{G}}$ approximately $10^{-8}$. For results on the bottom row this ratio was chosen as $0.230$, giving a demonstration how a conductive GB-phase would influence the currents. In the cases of a high $\sigma^\mathrm{GB}$ the influences of GB-phases on $|J|$ are small resulting in a straight current following the potential $\phi$ closely. Slight kinks appear as the currents cross the boundary perpendicular to the GB-phase-border and continue without any deflections to the GB-phase on the opposing side of the grain. This behavior is visible in all synthetic samples. In digitized samples the behavior diverted from this. As the currents flow around insular regions inside the grain, yet in the absence of these insular regions the previous behavior of straight forward flow occurs. Fig. \ref{fig:sigma_eff} (f$_4$) stands out in particular. As the sample with the largest GB-phase, the most conductive paths will be the ones containing less GB-phases and the current crosses the regions at thinner parts of the GB-phase.
%\newline
A different behavior is observed for the configuration containing resistive GB-phases, this is evident especially in the synthetic samples. In these synthetic cases strong pronounced kinks can be observed in the pathways of the current. These pathways cross GB-phases perpendicular and follow the shortest path throughout the more conductive grain. Furthermore the currents may show curved pathways in the grain interior. In the case of SEM-digitized samples the the currents are strongly deflected, due to the inhomogeneous microstructure preferred pathways occur, being able to majorly circumvent GB-phases, only crossing them in case no direct contact between grains exists. Coloring along the field lines indicates the magnitude of the current, with low currents occuring mostly directly at the GB-phase.\newline
Fig. \ref{fig:we} shows the calculated eddy currents, using eq. \ref{eq:we} for different reconstructions, both synthetic and SEM-digitized reconstructions are used. The corresponding $\WE$ of the microstructures are shown in Fig. \ref{fig:we}, also included are the eddy current losses assuming a low-frequency limit $W_\mathrm{E}^\mathrm{lf}$. As can be seen $\WE (\omega)$ follows $\omega$ linearly, though each microstructure exhibits its own specific slope. SEM-digitized samples and the synthetic reconstruction of sample B show comparable behavior, whereas the synthetic reconstruction of sample A shows significant deviation from the aforementioned three. For this sample $\WE$ exceeds the other three and this difference increases with increasing $\omega$. In this manner the the effective conductivity of sample A as shown in Fig. \ref{fig:sigma_eff} (c), where sample A shows higher conductivities over the whole range of $\sigma^{GB} /\sigma^{G}$. On the other hand both reconstructions of sample B show comparable results.
The variation of $\gl$ and $\gbw$ is shown in Fig. \ref{fig:we_var}, with (a) the variation of $\gl$ and (b) the variation of $\gbw$. In general the material with a lower silicon content exhibits higher losses compared to the one with a higher $\XSi$. This is due to the calculation of the conductivity, as an increased $\XSi$ leads to a lower conductivity of the material \cite{littmann1971iron}. 
Over all $\WE$ increases with $\gl$, as expected, as the decreasing volume fraction of the GB-phase provides less blocked paths and the currents can flow with less hindrance. An increase of $\gbw$ on the other hand leads to a decrease of $\WE$. This trend is due to $\sigma^\mathrm{GB} \ll\sigma^\mathrm{G}$. The increase of the overall volume of the resistive GB-phase leads to a reduced $\sigma^\mathrm{eff}$, and therefore reduced $\WE$.

\section{Conclusion}
In this work, we present a multiphysical approach to study the microstructure-property relations of the addively manufactured electrical steel. Digital microstructures were generated based on descriptors informed from experimental characterization and evaulated with hysteresis and eddy current losses employing Landau-Lifshitz and magneto-quasi-static model. This approach was validated to a certain extent by comparison to experimentally measured hysteresis of two distinct samples.

\begin{itemize}
     \item GB-phases turn to reduce the coercive fields. 
    \item The GB-phase influences the domain structure, introducing pinning sites into the system and restricting domains to singular grains. Preferred nucleation on the GB-phases could be observed. 
    \item Simulation results of synthetic reconstructions and SEM-digitized ones were in good agreement, suggesting the purely synthetic reconstructions can be sufficient to gather meaningful insights into the behavior of soft magnetic materials.
    \item We identified the trends of how key aspects of magnetic hysteresis change in response to different microstructural features. In particular, we demonstrated that an increase in the grain boundary thickness $\gbw$ can further decrease the coercive field. 
\end{itemize}

Furthermore we present a homogenization scheme for effective electric properties of the microstructure. Based on this we simulated the electric behavior of the microstructures and how electric currents respond to the introduction of GB-phases of different conductivities.
\begin{itemize}
     \item Results of the effective conductivities of various synthetic and digitized microstructures demonstrated that the GB-phase has significant influence on the effective conductivities.
    \item Through the spatial resolved numerical results, we reveal how the GB-phase influences the flow of currents in details. Therein we showed how GB-phases act as bottlenecks and strongly deflect the pathways of currents.
    \item We calculated further the eddy current losses of various microstructure samples. The increase of the average grain size $\gl$ leads to an increase in eddy current losses, whereas increasing the grain boundary thickness $\gbw$ decreases losses. The Gb-phase decreases the overall conductivity so increasing this even further can help reduce eddy currents in soft magnets.
\end{itemize}
 
Based on the current simulation results, further open questions are identified, which can be examined in the near future. For instance, the properties of the GB-phase are calculated as modified properties of $\fesi$, due to its distinct lamellar structure. An additional homogenization scheme could be developed to gain the knowledge about effective properties of these structures. Moreover, the reconstruction scheme of microstructure can be expanded, as the current ones cannot capture all features, such as isolated or insular GB-phases and pore phases. 

\section*{Data Availability}
	The authors declare that the data supporting the findings of this study are available within the paper. The simulation results and utilities are cured in the online dataset (DOI: \url{xx.xxxx/zenodo.xxxxxxx}). 

 \section*{Code Availability}
 Source codes of MOOSE-based application MFM-FM and related utilities are available and can be accessed via the online repository \url{https://git.rwth-aachen.de/tuda_mfm_packages/mfm-fm.git}. 

   \section*{Acknowledgements}
B.-X. Xu acknowledges the financial support of German Science Foundation (DFG) in the framework of the Collaborative Research Centre Transregio 361 (CRC-TRR 361, project number 492661287, sub-projects A05) and 270 (CRC-TRR 270, project number 405553726, sub-projects A06, B13, Z-INF), the Research Training Groups 2561 (GRK 2561, project number 413956820, sub-project A4), and the Priority Program 2256 (SPP 2256, project number 441153493). The authors also greatly appreciate the access to the Lichtenberg II High-Performance Computer (HPC) by the NHR Center NHR4CES@TUDa (funded by funded by the German Federal Ministry of Education and Research and the Hessian Ministry of Science and Research, Art and Culture), and High-performance computer HoreKa by the NHR Center NHR@KIT (funded by the German Federal Ministry of Education and Research and the Ministry of Science, Research and the Arts of Baden-Württemberg, partly funded by the German Research Foundation (DFG). The computating time on the HPC is granted by the NHR4CES Resource Allocation Board under the project ``special00007''.

\section{Competing Interests}
The authors declare no competing financial or non-financial interests.

\clearpage 

% \bibliography{reference}

% \clearpage

%%%%

..

%\begin{thebibliography}{00}
%\bibitem{}
%\end{thebibliography}
\end{document}